\documentclass[preprint,12pt,3p]{elsarticle}

\usepackage{framed,multirow}

\usepackage{amssymb}
\usepackage{amsmath}
\usepackage{latexsym}
\usepackage{graphicx}
\usepackage{dcolumn}
\usepackage{bm}
\usepackage{arydshln}

\usepackage{url}
\usepackage{xcolor}

\usepackage[colorlinks=true,citecolor=blue,linkcolor=blue,linktocpage=true,pagebackref=false]{hyperref}
\hypersetup{colorlinks=true,citecolor=blue,linkcolor=blue,filecolor=blue,urlcolor=blue}
\usepackage{color,soul}

\newcommand{\be}{\begin{equation}}
\newcommand{\ee}{  \end{equation}}
\newcommand{\ba}{\begin{eqnarray}}
\newcommand{\ea}{  \end{eqnarray}}

\newcommand\bra[1]{\left\langle {#1} \right|}
\newcommand\ket[1]{\left| {#1} \right\rangle}
\newcommand\braket[2]{\left\langle {#1} \middle| {#2} \right\rangle}
\newcommand\deriv[2]{\frac{\partial {#1}}{\partial {#2}}}

\graphicspath{{./figs/}}

\journal{Journal of Computational Physics}

\begin{document}

\begin{frontmatter}

    \title{A numerical method to efficiently calculate the transport properties of large systems: an algorithm optimized for sparse linear solvers}
    
    \author[1]{Tatiane P. {Santos}\corref{cor1}}
    \ead{tatiane.santos@if.uff.br}
    \cortext[cor1]{Corresponding author}
    \author[1]{Leandro R. F. {Lima}}
    \author[1]{Caio H. {Lewenkopf}}
    
    \address[1]{Instituto de F\'{\i}sica, Universidade Federal Fluminense, Av. Litor\^anea s/n, Niter\'oi,24210-346, Brazil}

\begin{abstract}
    We present a self-contained description of the wave-function matching (WFM) method to calculate electronic quantum transport properties of nanostructures using the Landauer-B\"uttiker approach. The method is based on a partition of the system between a central region (``conductor'') containing $N_S$ sites and an asymptotic region (``leads'') characterized by $N_P$ open channels. The two subsystems are linearly coupled and solved simultaneously using an efficient sparse linear solver. Invoking the sparsity of the Hamiltonian matrix representation of the central region, we show that the number of operations required by the WFM method in conductance calculations scales with $\sim N_S\times N_P$ for large $N_S$.
    \end{abstract}


\end{frontmatter}

\section{Introduction}
\label{sec:intro}

Advances in the fabrication of high-quality samples at the micro and nanoscale paved the way for the discovery of unusual electronic transport properties. As a consequence, the demand for numerical methods to realistically describe such systems on an atomistic/microscopic basis has dramatically increased. At low temperatures, experiments have reported electronic coherence lengths as long as tens of microns, typically larger than the characteristic sample size \cite{mayorov2011micrometer, banszerus2016ballistic, bandurin2017high, li2016quantum}. In such mesoscopic systems, the electronic transport is dominated by quantum interference. In this paper, we critically analyze the Wave Function Matching (WFM) method \cite{Groth2014,ando1991quantum,Khomyakov2005}, whose numerical implementations allow to efficiently compute the quantum transport properties of electrons in nanostructures, modeling realistic sample sizes and non-trivial geometries.

Quantum electronic transport in mesoscopic systems is usually described by the Landauer-B\"uttiker approach \cite{Datta1995}, that gives a simple relation between the conductance and the quantum transmission coefficients of a single-particle scattering problem. In other words, the problem is reduced to solving a Schr\"odinger equation for an open quantum system. We show in this paper that the WFM method is one of the most efficient numerical tools for this task. The latter introduces a partition between a central or scattering region (``conductor'') and the asymptotic one (``leads'' or terminals) and, by matching the corresponding wave function at the partition boundaries, gives the system scattering matrix $S$ \cite{Mello2004}.

Alternatively, transport properties in mesoscopic systems can be calculated using non-equilibrium Green's function (NEGF) techniques \cite{Datta1995}. This formalism is widely used due to its successful combination with Density Function Theory \cite{Taylor2001, brandbyge2002density, Ke2004, rocha2005towards, birner2007nextnano, bruzzone2014NanoTCADViDES,papior2017improvements}. At the single-particle level, NEGF is equivalent to the Landauer-B\"uttiker approach (see, for instance, Ref.~\cite{Hernandez2007}). The standard method to compute transport properties in large systems using NEGF is the Recursive Green's Function (RGF) method \cite{Thouless1981,MacKinnon1985,Lima2018}. The latter takes advantage of the sparsity of the system Hamiltonian to partition the scattering region into conveniently chosen small domains \cite{Wimmer2009,Cauley2011}. The corresponding Green's functions are recursively combined using the Dyson equation to obtain matrix elements of the full system Green's function that are relevant for transport calculations. The RGF method is robust, accurate, has a simple implementation, and has been widely used \cite{Baranger1988, Rotter2000,
Kazymyrenko2008, kuzmin2013OMEN, Libisch2012, Lewenkopf2013,Power2017}.

A recent open source implementation of the WFM method, the Kwant package \cite{Groth2014}, has significantly increased its usage. Kwant is developed under a user-friendly platform coded in Python and handles general-shaped scattering regions, multiple orbitals, and multi-probes \cite{wimmer2009quantum}. Furthermore, extensions to the Kwant package can be easily joined \cite{gaury2014numerical}. Kwant also explores the sparsity of the system Hamiltonian by using the MUMPS libraries, a forefront package for sparse linear algebra \cite{Amestoy2001}. Ref.~\cite{Groth2014} shows that Kwant significantly outperforms the RGF method in a wide range of applications.

In this paper, we show that the number of operations required by the WFM method to compute the conductance of a given system is much smaller than previously claimed \cite{Groth2014}. To explain this finding, we first give a self-contained presentation of the method -- whose documentation is scattered and scarce -- critically analyzing its main features. Next, we numerically study 
a number of systems to corroborate our analytical findings. 

The paper is organized as follows: In Sec.~\ref{sec:general_theory} we provide a short review of the relation between quantum transport and the scattering theory. Next, we adapt the theory for the tight-binding approximation and cast the scattering problem as the solution of a linear system. In Sec.~\ref{sec:WFM} we describe the WFM method and discuss its computational cost. In Sec.~\ref{sec:benchmarking} we benchmark the WFM method comparing its CPU time, memory usage and precision with a standard RGF implementation. Next, we present an application of the WFM method for a realistic-sized multi-probe graphene Hall bar system. In Sec.~\ref{sec:conclusion} we summarize our conclusions.

\section{Theoretical background}
\label{sec:general_theory}

The WFM method is suited to calculate the scattering properties of a system with arbitrary geometry and dimension $d = 1,2,$ or 3. It is aimed to describe the non-interacting on-shell scattering processes in mesoscopic samples or crystalline structures coupled to multiple terminals.

\begin{figure}[th!]
    \centering
    \includegraphics[width=8cm]{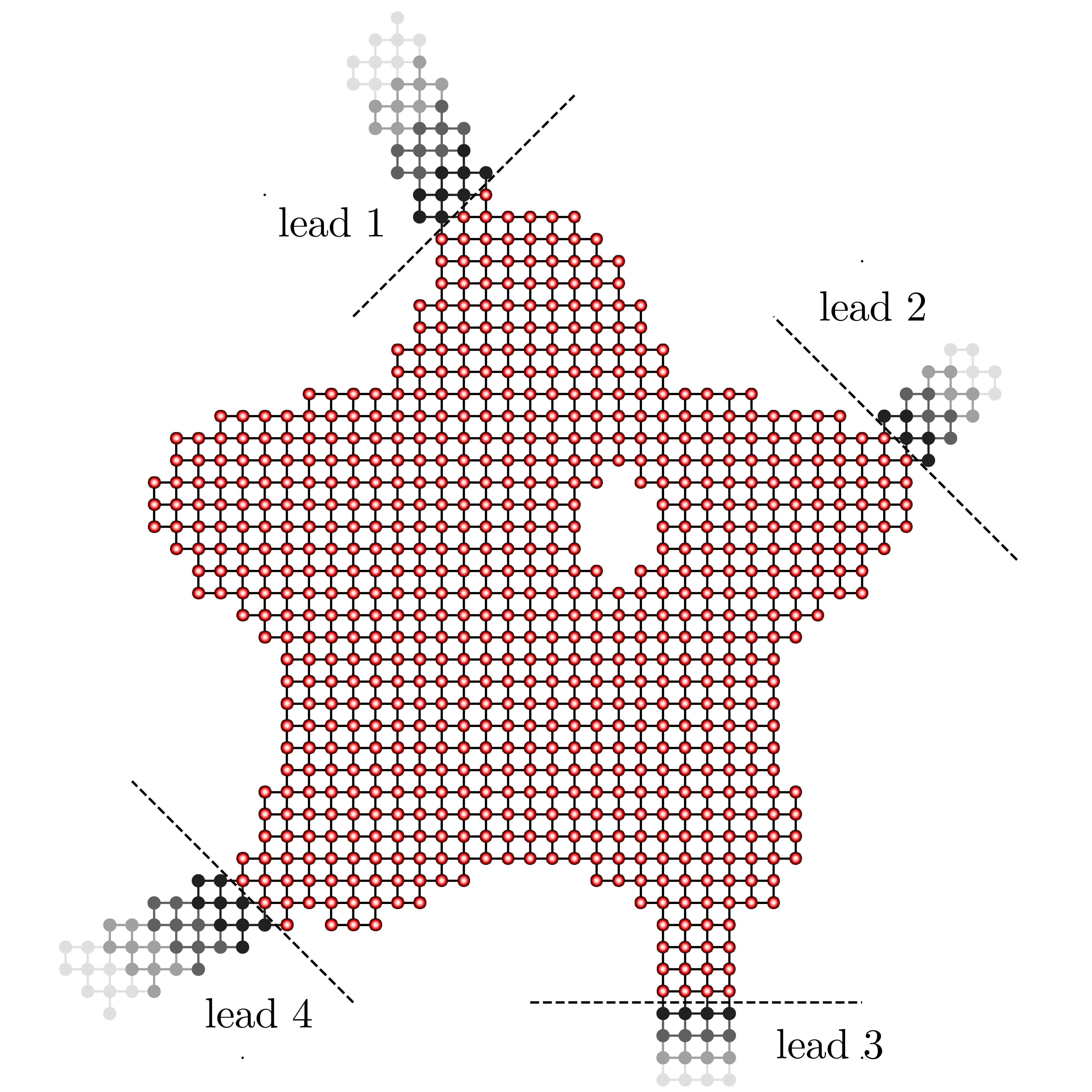}
    \caption{Illustration of a generic multi-terminal two-dimensional system. The dashed lines indicate the partition between the scattering and the asymptotic regions. The latter is modelled by or semi-infinite periodic lattices.}
    \label{fig:general_system}
    \end{figure}

Figure \ref{fig:general_system} illustrates a generic multi-terminal two-dimensional $(d=2)$ system. A central scattering region is coupled to electrodes represented by semi-infinite leads labeled by $\alpha = 1, \cdots, \Lambda$, where incoming and outgoing electrons propagate coherently. Due to the transverse confinement, the leads states are quantized in open modes (scattering channels) labeled by $n=1,\dots, N_{\alpha}$. The index $n$ labels both the transverse modes and the electron spin projection. The mesoscopic sample corresponds to the central or scattering region, while the leads are associated to the asymptotic domain.

Let us describe the system single-particle Hamiltonian by a tight-binding model. This approximation is suited to model both an atomistic system represented by a linear combination of atomic orbitals and a continuous system in a finite element representation \cite{Datta1995}. The system Hamiltonian is written as
\begin{equation}
H = \sum_{j, j'} H_{j, j'}| j \rangle \langle j'|,
\end{equation}
where the index $j=({\bm r}_i,\sigma)$ labels both the position in the lattice and the internal degrees of freedom $\sigma$ such as spin, atomic orbital, etc., of the state $\ket{j}$.

\subsection{Quantum transport and scattering theory}

The Schr\"{o}dinger equation of the scattering system reads
\begin{equation}\label{eq:schrodinger}
  H\ket{\Psi_{m}^{\pm}(E)}=E\ket{\Psi_{m}^{\pm}(E)},
\end{equation}
where $|\Psi_{m}^{+}(E)\rangle$ ($|\Psi_{m}^{-}(E)\rangle$) stands for the outgoing (incoming) scattering state at channel $m$. Here $m$ labels both $\alpha$ and $n$. The S-matrix is defined by the scattering amplitudes $\langle\Psi_{m}^{-}(E) | \Psi_{m'}^{+}(E)\rangle=S_{mm'}(E)\delta(E-E')$.

The scattering matrix $S$ can be formally written in terms of projection operators that decompose the Hilbert space in the partition described by Fig.~\ref{fig:general_system} \cite{Mahaux1969}. Let us assume, for instance, normal boundary conditions at the interface ${\cal B}$ between the scattering and asymptotic regions. One defines the projection operator 
\begin{equation}
Q = \sum_{\mu} \ket{\phi_\mu}\bra{\phi_\mu}
\end{equation}
in terms of the complete set of discrete orthonormal states $\braket{\phi_{\mu}}{\phi_{\mu'}}=\delta_{\mu\mu'}$ defined in the scattering (or central) region and obeying the boundary conditions at ${\cal B}$. In turn, at the asymptotic region, one defines
\begin{equation}
P = \sum_{m \in \alpha} \int d E\ket{\chi_{m}(E)}\bra{\chi_{m}(E)},
\end{equation}
where $\ket{\chi_{m}(E)}$ form a complete set of continuous orthogonal states, $\braket{\chi_{m}(E)}{\chi_{m'}(E')}=\delta_{mm'}\delta(E-E')$, defined in the asymptotic (or leads) region. Since the asymptotic region is not compact, the projection operator $P$ is continuous. By construction, $P$ and $Q$ span the system Hilbert space and, hence, $P + Q = 1$.

The system Hamiltonian is conveniently decomposed into three pieces
\begin{equation}
  H = H_{PP} + H_{QQ} +(H_{PQ}+H_{QP}),
\end{equation}
where we introduced the notation $AHB=H_{AB}$
\footnote{It is convenient to use as the channel basis in the asymptotic region the eigenfunctions of $H_{PP}$, namely, $H_{PP}\ket{\chi_m(E)}=E\ket{\chi_m(E)}$ with normal boundary conditions at ${\cal B}$.}.

The projection operators allow one to write Eq.~\eqref{eq:schrodinger} as a Lippmann-Schwinger equation, namely
\begin{align}
  P\ket{\Psi_{m}^{\pm}(E)} &= \ket{\chi_m(E)} + \frac{1}{E^{\pm}-H_{PP}}H_{QQ}Q\ket{\Psi_{m}^{\pm}(E)} \\
  Q\ket{\Psi_{m}^{\pm}(E)} &= \frac{1}{E^{\pm}-H_{QQ}}H_{QP}P\ket{\Psi_{m}^{\pm}(E)},
\end{align}
where $E^\pm = E \pm i \eta$, with $\eta$ an infinitesimal positive number. After some algebra, one writes the $S$-matrix as \cite{Mahaux1969,Mello2004}
\begin{equation}
S_{mm'}(E)=\delta_{mm'}-2i\pi \rho_m^{1/2}(E)\sum_{\mu\mu'}[H_{PQ}]_{m\mu}\left[\frac{1}{E-H_{QQ}-\Sigma^{+}(E)}\right]_{\mu\mu'}[H_{QP}]_{\mu'm'} \rho_{m'}^{1/2}(E),
\label{smatrix0}
\end{equation}
where $\Sigma^{\pm}(E)=H_{QP}(E^{\pm}-H_{PP})^{-1}H_{PQ}$ is the embedding self-energy, which accounts for coupling the to the continuum and describes the resonance processes, while $\rho_m(E)$ stands for the electronic density of states at the channel $m$. Here we explicitly neglect direct tunneling processes between different electrodes \cite{Mahaux1969, Mello2004}. This approximation is accurate provided the central region is sufficiently large to prevent direct tunneling processes across the system. This condition is true for most mesoscopic systems, except for small molecular junctions (for more details, see, for instance, Ref.~\cite{DiVentra2008}).

The Landauer-B\"{u}ttiker theory \cite{Datta1995} relates the linear conductance of an electronic sample to the transmission probability as
\begin{equation}\label{eq:landauer-buttiker}
  {\cal G}_{\alpha\beta}=\frac{e^2}{h}\int_{-\infty}^{\infty}d E \left(-\deriv{f}{E}\right) T_{\alpha\beta}(E),
\end{equation}
where $f(E)=\left[1+e^{(E-\mu)/kT}\right]^{-1}$ is the Fermi-Dirac distribution with $\mu$ and $T$ giving the equilibrium chemical potential and temperature of the reservoirs
\footnote{In general, the reservoirs have different chemical potentials and temperatures, thus, $f_\alpha(E)=[1+e^{(E-\mu_{\alpha})/kT_{\alpha}}]^{-1}$. For simplicity we take all temperatures equal to $T$ and since we restrict ourselves to linear response, the small differences between $\mu_{\alpha}$ and the equilibrium $\mu$ lead to Eq.~\eqref{eq:landauer-buttiker}.}
. The transmission $T_{\alpha\beta}(E)$ is given by 
\begin{align}
	T_{\alpha\beta}(E) = \sum_{\substack{n\in\alpha \\ m\in\beta}} \left|S_{nm}(E)\right|^2,
\end{align}
where $S_{nm}(E)$ is given by Eq.~\eqref{smatrix0}. The WFM method also gives local properties such as local currents and the local density of states (LDOS), as discussed in Sec.~\ref{sec:benchmarking}.

\subsection{The scattering problem in tight-binding approximation}
\label{sec:tight-binding}

Let us now write the system Hamiltonian in a suitable form to implement the WFM method. For the sake of simplicity, 
we discuss in detail the two-terminal case and, at the end, we generalize the results to the multi-terminal case.

\begin{figure}[!htb]
    \centering
     \includegraphics[width=16cm]{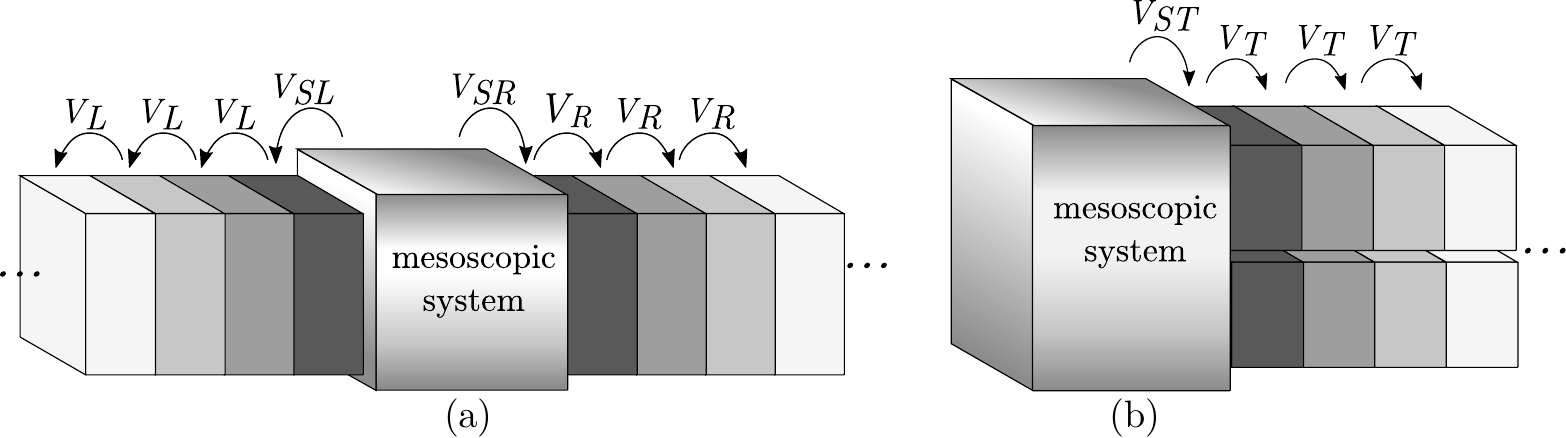}
      \caption{
      (a) Sketch of a mesoscopic system (S) coupled to Left (L) and Right (R) semi-infinite leads with periodic lattice structure.
      (b) Equivalent system with the L and R terminal (in general, $\alpha= 1, \cdots \Lambda$) mapped into a single-lead.
      }
      \label{fig:mesoscopicsystem}
   \end{figure}

Let us consider a mesoscopic system attached to semi-infinite leads, $\alpha=R,L$, as illustrated by 
Fig.~\ref{fig:mesoscopicsystem}a. Following the partition operators presented in the previous section, we introduce 
the standard matrix representation: 
(i) $H_{QQ} \leftrightarrow H_S$ for the scattering region Hamiltonian;
(ii) $H_{PP} \leftrightarrow H_{L}+H_{R}$, for the leads Hamiltonian;
(iii) $H_{QP} \leftrightarrow V_{SL}+V_{SR}$, for the coupling term connecting the mesoscopic system to the leads.

The full Hamiltonian is written in a block matrix form as
\begin{equation}
H = \begin{pmatrix}
H_L & V_{LS} & 0\\
V_{SL} & H_S & V_{SR}\\
0 & V_{RS} & H_R
\end{pmatrix}.
\end{equation}
$H_L$ and $H_R$ can be written in the block-diagonal structure
\begin{equation}
H_{L} = \begin{pmatrix}
\ddots & \ddots & & \\
\ddots & H_l & V_l & \\
& V_l^{\dagger} & H_l & V_l \\
& & V_l^{\dagger} & H_l
\end{pmatrix}\qquad
H_{R} = \begin{pmatrix}
H_r & V_r^{\dagger} & & \\
V_r & H_r & V_r^{\dagger} & \\
& V_r & H_r & \ddots \\
& & \ddots& \ddots
\end{pmatrix},
\end{equation}
where $H_{l(r)}$ stands for suitable $L(R)$-lead unit cell Hamiltonian of dimension $M_{L(R)}$ (represented by boxes 
in Fig.~\ref{fig:mesoscopicsystem}). $V_{l(r)}$ are the hopping matrices between nearest-neighboring unit-cells and the 
unwritten matrix elements are identically zero.

It is advantageous to use the structure of the leads matrices $H_L$ and $H_{R}$ to group them into an effective single-lead with disjoint sections. The  rearranged layout is depicted in Fig. \ref{fig:mesoscopicsystem}b. The modified $H$ reads
\begin{equation}\label{eigenproblem2}
{\small
\setlength{\arraycolsep}{2.5pt} 
\medmuskip = 1mu
H=\left(\begin{array}{c:c c:c c c}
H_S-E & V_{SL}^{\dagger} & V_{SR}^{\dagger} & & & \\
\hdashline
V_{SL} & H_l-E & 0 & V_l^{\dagger} & 0 & \\
V_{SR} & 0 & H_r-E & 0 & V_{r}^{\dagger} &  \\
\hdashline
& V_{l} & 0 & H_l-E & 0 & \\
& 0 & V_{r} & 0 & H_r-E & \ddots \\
& & & & \ddots & \ddots
\end{array}\right).}
\end{equation}

The effective lead, hereafter denoted by $T$, compacts the eigenvalue problem to a single semi-infinite partition, namely
\begin{equation}\label{eigenproblem3}
\begin{pmatrix}
H_S-E & V_{TS}^{\dagger}& & \\
V_{TS} & H_T-E & V_T^{\dagger} & & \\
& V_T & H_T-E & \ddots\\
& & \ddots & \ddots
\end{pmatrix}
\begin{pmatrix}
\psi_S \\
\psi_{0} \\
\psi_{1} \\
\vdots
\end{pmatrix} =
\begin{pmatrix}
0 \\
0 \\
0 \\
\vdots
\end{pmatrix},
\end{equation}

\noindent where $\psi_S$ corresponds to the scattering wave function at the central region and $\psi_n$, to the lead wave function at the $n$-th slice, with $n=0,0,2,\dots$ (see Fig. \ref{fig:mesoscopicsystem}). The generalization to a multi-terminal setup is straightforward. In this case, $H_T$ accounts for all 
$H_{\alpha}$, with $\alpha=1,\dots,\Lambda$ and has dimension $M_T=\sum_{\alpha=1}^{\Lambda}M_{\alpha}$.

\section{The wave function matching method}
\label{sec:WFM}

Let us now solve Eq.~(\ref{eigenproblem3}). For that purpose we introduce the eigenmode basis $\phi_n$:
\begin{align}\label{lead_equation}
V_{T}\phi_{n-1} + (H_T-E)\phi_{n} + V_T^{\dagger}\phi_{n+1} = 0,
\end{align}
which corresponds to rows of Eq.~(\ref{eigenproblem3}) far from the scattering region. Due to translational symmetry, one can use Bloch's theorem to conveniently write $\phi_{n}$ as 
\begin{align}\label{chi_lambda}
\phi_{n} = \chi \lambda^{n},
\end{align}
where $\chi$ is the lead unit cell eigenfunction (independent of $n$) and $\lambda$ is a complex constant. Hence,
\begin{align}
\label{QEP}
V_{T}\chi + (H_T-E)\chi\lambda + V_T^{\dagger}\chi\lambda^2 = 0.
\end{align}
The standard procedure to solve this quadratic eigenvalue problem (QEP) in $\lambda$ is to introduce an auxiliary vector 
\begin{align}\label{chiprime}
	\chi' \equiv \lambda^{-1} V_T \chi
\end{align}
and to linearize Eq.~(\ref{QEP})\cite{Golub1996} as 
\begin{align}
\begin{pmatrix}
H_T-E & \hat{1} \\
V_T & \hat{0}
\end{pmatrix}
\begin{pmatrix}
\chi\\
\chi'
\end{pmatrix}
=
\lambda\begin{pmatrix}
-V_{T}^{\dagger} & \hat{0} \\
\hat{0} & \hat{1}
\end{pmatrix}
\begin{pmatrix}
\chi \\
\chi'
\end{pmatrix}
.
\label{GEP}
\end{align}

\noindent The advantage of casting Eq.~(\ref{QEP}) as a Generalized eigenvalue problem (GEP) is that one can calculate  the eigenvalue $\lambda$, which is associated to the crystal momentum $k$ (using $\lambda=e^{ika_\alpha}$, where $a_{\alpha}$ is the $\alpha$-lead lattice constant), and the eigenvector $\chi$ as a function of the electronic energy $E$. The QEP is translated into a linear problem at the expense of doubling the equation dimension. Hence, the number of eigenvalues is twice the rank $M_T$ of the matrices $V_T$ and $H_T$. 

One can solve the GEP in Eq.~(\ref{GEP}) by means of well-known numerical algorithms \cite{Golub1996,Rungger2008, Anderson1990,Anderson1999}. Given an electronic energy $E$, we calculate the eigenvectors $(\chi_p, {\chi'}_p)$ and the corresponding eigenvalues $\lambda_p$, where $p=1,\cdots,2M_T$.

\begin{figure}[th!]
	\centering
	\includegraphics[width=10cm]{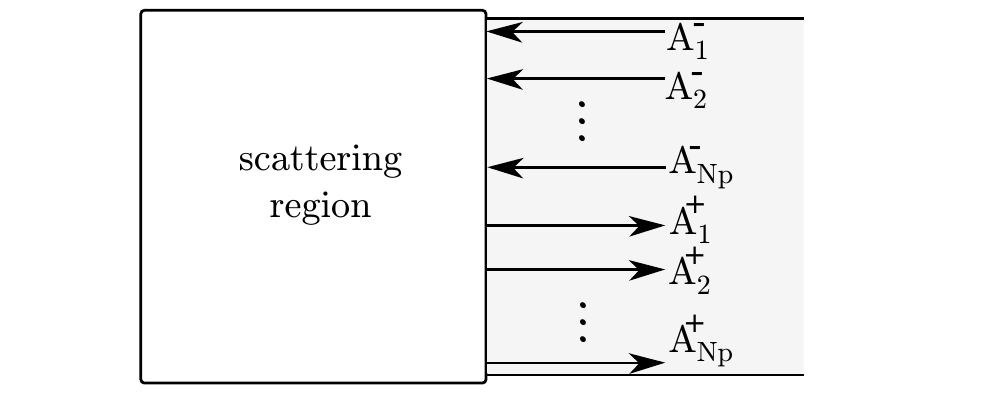}
	\caption{\small Multi-mode representation of the scattering process of a single-lead with 
	$M_T=\sum_{\alpha=1}^{\Lambda}M_{\alpha}$ modes. The sign $-$ ($+$) indicates incoming (outgoing) 
	modes with amplitude $A_p^-$ ($A_p^+$), where $p=1,2,\cdots,N_P$.}
	\label{multi_mode}
\end{figure}

We can infer from the scattering problem that the $2M_T$ solutions correspond to $M_T$ incoming modes and $M_T$ outgoing modes, as depicted in Fig.~\ref{multi_mode}. Since the terminals are uncoupled, the eigenstate $\chi_p$ has a block structure 
\begin{align}
	\chi_p = \left(\cdots, 0, \chi_p^\alpha, 0, \cdots\right),
\label{chiblock}
\end{align}
where each block $\chi_p^\alpha$ describes the eigenstate of the $\alpha$-lead with eigenvalue $\lambda_p$ for $p=1,\cdots,M_\alpha$ and $\alpha=1,\cdots,\Lambda$.

The modes can be propagating $|\lambda_{p}| = 1$ or evanescent $|\lambda_{p}|< 1$  ($|\lambda_{p}| > 1$ gives a non-physical behavior). The probability current for the $p$-th propagating mode reads \cite{Zwierzycki2008}
\begin{align}
	j_p 	&= -\frac{2}{\hbar}\text{Im}\left(\lambda_p\chi_p^\dagger V_T\chi_p\right).
	\label{current}
\end{align}
The incoming modes correspond to $j_p>0$ and the outgoing ones to $j_p<0$. We label those two sets of solutions as $\lambda_p^{\pm},\chi_p^{\pm}$ for $p=1,\dots,N_P$, where $\pm$ indicates the corresponding current direction and $N_P \leq M_T$ \label{normalize notation} is the number of incoming/outgoing propagating channels at the electronic energy $E$. See Fig.~\ref{multi_mode}. Since we are interested in transmission coefficients, we restrict ourselves to the analysis of the propagating modes. The evanescent modes ($j_p=0$) can be treated straightforwardly as a generalization of this method. 

Using the sets $\chi_p^{\pm}$ as basis, we write the wave functions $\psi_n$ as
\begin{align}\label{basis}
\psi_n = \sum_{q=1}^{N_P}A_{q}^{-}\chi_{q}^{-}({\lambda_{q}^{-}})^n + \sum_{p=1}^{N_P}A_{p}^{+}\chi_{p}^{+}({\lambda_{p}^{+}})^n,
\end{align}
where $n=0,1,\cdots$, and $A_p^\pm$ are unknown amplitudes.

One defines the scattering matrix $\tilde S$ that relates incoming with outgoing amplitudes as
\begin{align}
\begin{pmatrix}
A_{1}^{+}\\
A_{2}^{+}\\
\vdots\\
A_{N_P}^{+}
\end{pmatrix} = \tilde S
\begin{pmatrix}
A_{1}^{-}\\
A_{2}^{-}\\
\vdots\\
A_{N_P}^{-}
\end{pmatrix}
. \label{ASB}
\end{align}
Since the eigenchannel basis used by the WFM method is not normalized as the one introduced in Sec.~\ref{sec:general_theory}, the matrix $\tilde S$ does not preserve the flux. As we discuss below, $S$ is obtained from $\tilde S$ by a simple relation.

To calculate the S-matrix, we consider the scattering process of a single incoming mode $q$, namely
\begin{align}\label{injection1}
	\psi_{nq}  
  & = \chi_{q}^-({\lambda_{q}^-})^n+\sum_{p=1}^{N_P}\chi_{p}^+({\lambda_{p}^+})^n\tilde S_{pq}.
\end{align}
The corresponding S-matrix can be obtained by solving the first two lines of Eq.~(\ref{eigenproblem3})
\begin{align}
 (H_S-E)\psi_{Sq} + V_{TS}^{\dagger}\psi_{0q} = 0,										
 \label{final_problema} \\
 V_{TS}\psi_{Sq} + (H_T-E)\psi_{0q} + V_{T}^{\dagger}\psi_{1q} = 0,	\label{final_problemb}
\end{align}
where $\psi_{Sq}$ is the scattering region wave function upon injection from mode $q$. Substituting Eq.~(\ref{injection1}) into Eq.~(\ref{final_problemb}) and recalling that the basis functions $\chi^{\pm}_p$ satisfy Eq.~(\ref{QEP}), we find 
\begin{align}
V_{TS}\psi_{Sq} = V_T\psi_{-1q},
\label{eigenproblem02}
\end{align}
where $\psi_{-1q}$ is also given by Eq.~(\ref{injection1}). Note, however, that $\psi_{-1q}$ has no physical meaning, since in Eq.~(\ref{eigenproblem3}) there is no slice defined for $n=-1$. Here, $\psi_{-1q}$ is an auxiliary mathematical quantity designed to represent the contributions of the terms including $\psi_{0q}$ and $\psi_{1q}$ in Eq.~(\ref{final_problemb}). 

Applying the definition of $\chi'$, Eq.~(\ref{chiprime}), to each propagating mode as 
\begin{align}
	{\chi'}^{\pm}_q=\left(\lambda_q^{\pm}\right)^{-1} V_T \chi_q^{\pm},
\label{chiprimep}
\end{align}
Eq.~(\ref{eigenproblem02}) becomes
\begin{align}
V_{TS}\psi_{Sq} = {\chi'}_{q}^- + \sum_{p=1}^{N_P}{\chi'}_{p}^+ \tilde S_{pq} = {\chi'}_{q}^- + {\chi'}^+ \tilde S_q,
\label{eigenproblem03}
\end{align}
where ${\chi'}^\pm \equiv ({\chi'}^\pm_1, {\chi'}^\pm_2, \cdots, {\chi'}^\pm_{N_P})$ with dimension $M_T\times N_P$ and $\tilde S_q$ is the column $q$ of the S-matrix with dimension $N_P\times 1$. Analogously, using Eq.~(\ref{injection1}) we write $\psi_{0q}$ as
\begin{align}
\psi_{0q} = \chi_{q}^- + \chi^+ \tilde S_q.
\label{eigenproblem04}
\end{align}
The linear system composed by Eqs.~(\ref{final_problema}) and (\ref{eigenproblem03}) reads
\begin{align}\label{finalq}
\begin{pmatrix}
H_S-E 	& V_{TS}^{\dagger}\chi^+\\
V_{TS} 	& - {\chi'}^{+}
\end{pmatrix}
\begin{pmatrix}
\psi_{Sq} \\ \tilde S_q
\end{pmatrix} =
\begin{pmatrix}
-V_{TS}^{\dagger}\chi^{-}_q \\
{\chi'}^{-}_q
\end{pmatrix}
.
\end{align}
Let us now generalize Eq.~(\ref{finalq}) to account for different $q$-modes
\begin{align}\label{finalall}
\begin{pmatrix}
H_S-E 	& V_{TS}^{\dagger}\chi^+\\
V_{TS} 	& - {\chi'}^{+}
\end{pmatrix}
\begin{pmatrix}
\psi_{S1} & \psi_{S2} & \cdots & \psi_{SN_P}  \\ 
\tilde S_1 & \tilde S_2 & \cdots & \tilde S_{N_P} 
\end{pmatrix} = 
\begin{pmatrix}
-V_{TS}^{\dagger}\chi^{-}_1 &-V_{TS}^{\dagger}\chi^{-}_2 & \cdots &-V_{TS}^{\dagger}\chi^{-}_{N_P}  \\
{\chi'}^{-}_1 &  {\chi'}^{-}_2 & \cdots & {\chi'}^{-}_{N_P} 
\end{pmatrix}
.
\end{align}
We cast this result into the compact form
\begin{align}\label{finalfull}
\begin{pmatrix}
H_S-E 	& V_{TS}^{\dagger}\chi^+\\
V_{TS} 	& - {\chi'}^{+}
\end{pmatrix}
\begin{pmatrix}
\psi_S \\ \tilde S
\end{pmatrix} =
\begin{pmatrix}
-V_{TS}^{\dagger}\chi^{-} \\
{\chi'}^{-}
\end{pmatrix}
,
\end{align}
where $\tilde S$ is the full S-matrix and $\psi_S=\begin{pmatrix} \psi_{S1} & \psi_{S2} & \cdots & \psi_{SN_P} \end{pmatrix}$ is the wave function of the scattering region. The S-matrix has dimension $N_P\times N_P$ while $\psi_S$ has dimension $N_S \times N_P$, since it is defined for all the $N_S$ sites in the central region upon injection from all the $N_P$ channels. 

Hence, the solution of Eq.~(\ref{finalfull}) has a computational cost that depends on the number of propagating channels $N_P$ at the electronic energy $E$. Due to the sparsity of $H_S$, we infer that CPU time required to compute a given system conductance scales as $N_S \times N_P$. In Sec.~\ref{sec:benchmarking} we numerically verify that the WMF method indeed follows this prediction.

Note that Eq.~(\ref{finalfull}) involves representations in different spaces, while the scattering wave function is given in the tight-binding basis, the S-matrix is expressed in eigenmode basis. The matrices $\chi^{\pm}$ give a connection between theses two basis \cite{Ferry1997}. For a sufficiently large system, $H_S$ and $V_{TS}$ are sparse matrices making the problem appropriate to the sparse solvers.

\subsection{Connection to Green's functions}

The coupling with leads gives a finite line-width to the resonances in the scattering region via a so-called self-energy. In the non-equilibrium Green's functions formalism (NEGF) (see, for instance, Refs. \cite{Datta1995,DiVentra2008}) the embedding self-energy modifies the scattering region Hamiltonian as $H_S \rightarrow H_S+\Sigma$. In what follows we demonstrate that $\Sigma$ can be calculated from the presented equations.

Let us define the dual space states $\tilde\chi_p^{\pm}$, where 
\begin{align}
		\left(\tilde\chi_p^{\pm}\right)^\dagger\chi_{p'}^{\pm} = \delta_{pp'},
	\label{dualvectors}
\end{align}
and identify the first and the second terms on the RHS of Eq.~(\ref{injection1}) with 
\begin{align}
	\psi_{nq-}  \equiv ({\lambda_{p}^-})^n\chi_{q}^-,\quad \mbox{ and } \quad
	\psi_{nq+} \equiv \sum_{p=1}^{N_P}({\lambda_{p}^+})^n\chi_{p}^+\tilde S_{pq}. 	
\label{psinq+}
\end{align}
Introducing the translation operator $F_{\pm}$ \cite{Khomyakov2005} 
\begin{align}
	F_{\pm} \psi_{nq\pm} = \psi_{n+1,q\pm}, 
	\label{Fdefinition}
\end{align}
where
\begin{align}
	F_{\pm} \equiv  \sum_p^{N_P} \lambda_p^{\pm}\chi_p^{\pm}\left(\tilde\chi_{p}^{\pm}\right)^\dagger,
\end{align}
one can write $\psi_{0q}$ and $\psi_{1q}$, respectively, as
\begin{align}
	\psi_{0q} &= \psi_{0q-}+\psi_{0q+}, \label{psi0pm}\\
	\psi_{1q} &= F_-\psi_{0q-} + F_+\psi_{0q+} 
	= (F_--F_+)\chi_{q}^- + F_+\psi_{0q}. \label{psi1pm}
\end{align}
Substituting Eq.~(\ref{psi1pm}) into Eq.~(\ref{final_problemb}) and solving Eq.~(\ref{final_problema}) for $\psi_{Sq}$ we find
\begin{align}
 (E-H_S-\Sigma)\psi_{Sq} = Q_q^-, 
\label{psisqsource}
\end{align}
where
\begin{align}
	Q_q^- &\equiv V_{TS}^{\dagger} G_T V_{T}^{\dagger}(F_--F_+) \chi_{q}^-
\label{source}
\end{align}
is a source term dependent of which channel $q$ is injecting,
\begin{align}
	\Sigma = V_{TS}^{\dagger} G_T V_{TS}
\end{align}
is the embedding self-energy, and
\begin{align}
	G_T = \left(E-H_T-V_{T}^{\dagger}F_+\right)^{-1}
	\label{surfaceGF}
\end{align}
is the surface Green's function of the semi-infinite leads. Since Eq.~(\ref{surfaceGF}) involves outgoing states $F_+$, $G_T$ and $\Sigma$ correspond to retarded Green's function and self-energy, respectively \cite{Datta1995}.

We stress that both $\Sigma$ and $G_T$ are independent of $q$ and contain information about all the propagating modes at the energy $E$. In \ref{app:surfaceGF} we show that Eq.~(\ref{surfaceGF}) reproduces the exact analytical surface Green's function of a 1D chain.

Notice that we can solve Eq.~(\ref{psisqsource}) for $\psi_{Sq}$ as
\begin{align}
	\psi_{Sq} = G_SQ_q^-,
	\label{psiSqGF}
\end{align}
where $G_S$ is the scattering region Green's function given by
\begin{align}
	G_S \equiv \left(E-H_S-\Sigma\right)^{-1}.
	\label{GS}
\end{align}
Thus, knowing the full Green's function matrix $G_S$, we can calculate $\psi_{Sq}$ for any $q$ using Eq.~(\ref{psiSqGF}).

With the help of the dual vector $\tilde\chi_p^+$ defined in Eq.~(\ref{dualvectors}) and the definition of $\psi_{0q+}$ given by Eq.~(\ref{psinq+}), we calculate the amplitudes $\tilde S_{pq}$ as
\begin{align}
	\tilde S_{pq} = \left(\tilde\chi_p^+\right)^\dagger \psi_{0q+}.
	\label{spq}
\end{align}
The outgoing wave function $\psi_{0q+}$ is a superposition of states $\chi_p^+$ with amplitudes $\tilde S_{pq}$. Those states carry a probability current
\begin{align}
	j_{pq} = j_p\left|\tilde S_{pq}\right|^2.
	\label{currentpq}
\end{align}
Here $j_{pq}$ depends on the injecting mode $q$ and $j_p$ is given by Eq.~(\ref{current}).

The transport coefficients $P_{pq}$ defined as the ratio between the incoming probability current $j_q$ and the outgoing probability current $j_{pq}$ at mode $p$ reads
\begin{align}
	P_{pq} &= \frac{j_{pq}}{j_q} = \left|\sqrt{\frac{j_p}{j_q}}\tilde S_{pq}\right|^2= \left|S_{pq}\right|^2,
	\label{12345}
\end{align}
where we defined the scattering amplitudes $S_{pq}$ as \cite{Datta1995}
\begin{align}
	S_{pq} \equiv \sqrt{\frac{j_p}{j_q}}\tilde S_{pq}.
	\label{scatteringamplitudes}
\end{align}
\noindent where $S$ is unitary and conserves the current probability \cite{Datta1995}.

\subsection{Generalized Fisher-Lee expression for the transmission amplitudes}
\label{app:fisherlee}

Let us use the WFM method elements introduced above to derive the relation between the transmission amplitudes as 
functions of the scattering region Green's functions.

First we write $\psi_{0q+}$ in the RHS of Eq.~(\ref{psi0pm}) as a function of the scattering region wave function $\psi_{Sq}$ using Eqs.~(\ref{psi1pm}) and Eq.~(\ref{final_problemb}), namely
\begin{align}
	\psi_{0q+} &= G_TV_{TS}\psi_{Sq} + \left[G_TV_T^\dagger\left(F_--F_+\right)-1\right]\chi_q^-.
	\label{psi0q+gf}
\end{align}
Hence, the scattering amplitude $S_{pq} = \sqrt{{j_p}/{j_q}}\tilde S_{pq}$  reads
\begin{align}
	S_{pq} = \sqrt{\frac{j_p}{j_q}}\left(\tilde\chi_p^+\right)^\dagger  G_TV_{TS}G_SV_{ST} G_T V_{T}^{\dagger}(F_--F_+) \chi_{q}^- 
	+ \left(\tilde\chi_p^+\right)^\dagger \left[G_TV_T^\dagger\left(F_--F_+\right)-1\right] \chi_q^-.
	\label{fisherlee0}
\end{align}
Here we used Eqs.~(\ref{source}) and (\ref{psiSqGF}) to substitute the dependence on $\psi_{Sq}$ by a dependence on the scattering region Green's function $G_S$.

We assume that the modes $q$ and $p$ belong to different leads $\alpha$ and $\beta$, respectively. Due to the block structure of Eq.~(\ref{chiblock}) and to the absence of coupling between the leads, the matrices $G_T$, $V_T^\dagger$ and $\left(F_--F_+\right)$ are block diagonal in the leads subspace. The two-contacts Hamiltonian in Eq.~(\ref{eigenproblem2}) illustrates the diagonal block structure of $V_T^\dagger$, for instance. Thus, the second term in Eq.~(\ref{fisherlee0}) identically vanishes.

In this case, the scattering amplitude in Eq.~(\ref{fisherlee0}) becomes 
\begin{align}
	 t_{pq}^{\beta\alpha} = \sqrt{\frac{j_p}{j_q}}\left(\tilde\chi_p^+\right)^\dagger  G_TV_{TS}G_SV_{ST} G_T V_{T}^{\dagger}(F_--F_+) &\chi_{q}^-,
	\label{fisherlee1}
\end{align}
where $t_{pq}^{\beta\alpha}$ is the current-normalized transmission amplitude for the scattering from mode $q$ in the lead $\alpha$ to the mode $p$ in the lead $\beta$.

Although one can calculate the transmission coefficients by means of $G_S$ from Eq.~(\ref{fisherlee1}), only few Green's functions matrix elements, such as the elements connecting sites belonging to the interface with the leads, are required to compute the transmission (see, for instance Ref. \cite{Lima2018}). Therefore, a simplification of Eq.~(\ref{fisherlee1}) is desirable. For that purpose we use a sub-block division of the scattering region similar to the one used in Ref.~\cite{Lima2018}.

We divide the scattering region into $\Lambda+1$ blocks, where $C$ is the a central block, which has no connection with the leads, and $\alpha$ represents the $\alpha$-interface, which is connected to $C$ and only to the lead $\alpha$, where $\alpha=1,\cdots,\Lambda$. In this picture, $G_S$ and $V_{TS}$ read
\begin{align}
	G_S &=
	\left(\begin{array}{ccccc}
		[G_S]_{CC}				& [G_S]_{C1} 				& \cdots 	& [G_S]_{C\Lambda} \\   
		\left[G_S\right]_{1C}				& [G_S]_{11} 				& \cdots 	& [G_S]_{1\Lambda} \\  
		\vdots						& \vdots						&  				& \vdots 						\\ 
		\left[G_S\right]_{\Lambda C} & [G_S]_{\Lambda 1} & \cdots 	& [G_S]_{\Lambda\Lambda} 
	\end{array}\right),
	\\
	V_{TS} &=
	\left(\begin{array}{cccccc}
		0 			& V_{11}		&0					& \cdots	& 0			\\ 
		0 			& 0 				& V_{22}		& \cdots	& 0			\\ 
		\vdots	& \vdots		& \vdots		&					&\vdots	\\ 
		0 			& 0 				& 0 				& \cdots 	& V_{\Lambda \Lambda}
	\end{array}\right).
\end{align}
Since $G_TV_T^\dagger\left(F_--F_+\right)$ is diagonal, where $[G_T]_{\alpha\beta} = \delta_{\alpha\beta} G_\alpha V_\alpha^\dagger\left(F_-^\alpha-F_+^\alpha\right)$ and $F_\pm^\alpha \equiv \sum_{p\in \alpha}^{N_P} \lambda_p^{\pm}\chi_p^{\pm \alpha}\left(\tilde\chi_{p}^{\pm \alpha}\right)^\dagger$, and the states $\chi_{q}^-$ and $\tilde\chi_p^+$ have different non-vanishing blocks given by Eq.~(\ref{chiblock}), we find
\begin{align}
	t_{pq}^{\beta\alpha} = \sqrt{\frac{j_p}{j_q}}\left(\tilde\chi_p^{+\beta}\right)^\dagger  G_\beta V_{\beta\beta} \left[G_S\right]_{\beta\alpha} V_{\alpha\alpha} G_\alpha V_\alpha^\dagger(F_-^\alpha-F_+^\alpha) &\chi_q^{-\alpha},
	\label{fisherlee}
\end{align}
which is the generalized Fisher-Lee expression \cite{Datta1995}.

\section{Benchmark and Application}
\label{sec:benchmarking}

Let us now demonstrate the efficiency of the sparse solvers associated with the WFM method implemented in the Kwant package. To this end, we compare the processing time and memory usage of the WFM method with the standard RGF approach for a two-dimensional model system as a function of its size and aspect ratio. We conclude this section by discussing an application of the WFM method, namely, the calculation of longitudinal and transverse resistance of a realistic-sized graphene Hall bar.

As mentioned in the introduction, nowadays the RGF method is one of the most standard technique to compute the conductance of nanoscale systems. This method is designed to compute only the system full Green's function matrix elements related to transport properties \cite{Datta1995}. For that purpose, the system is divided into partitions. The computational time necessary to calculate the transmission scales with the number of partitions times the cube of the typical number of sites within the partitions.

We recall that Ref.~\cite{Groth2014} draws conclusions by comparing the performance of the RGF and WFM methods for a square lattice system with $L\times L$ sites as a function of $L$. The authors \cite{Groth2014} find that the CPU time required to compute the conductance using the RGF method scales with $L^4$, while the WFM implementation in Kwant scales with $L^3$. Here we explore more diverse situations to numerically verify that the WFM method is more efficient than $L^3$, as discussed in Sec.~\ref{sec:WFM}.

\begin{figure}
 \centering
   \includegraphics[width=10cm]{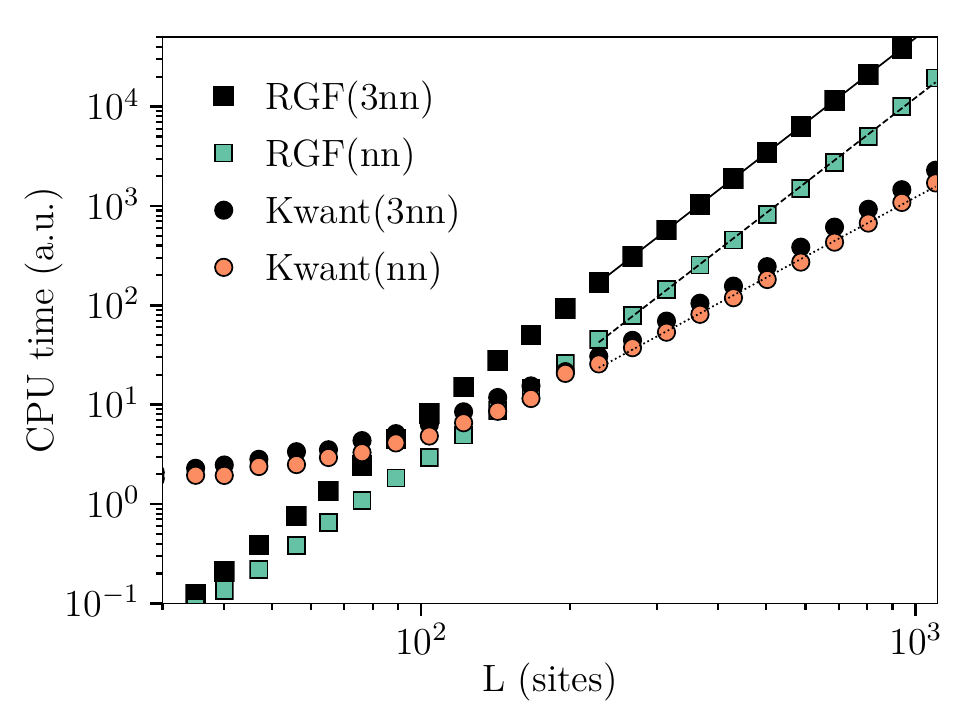}
 \caption{CPU time as a function of the side of a $L\times L$ system. The lines correspond to the best $a W^b$ fit. For Kwant (nn) $b \approx 2.7$ (dotted line), for RGF (nn) $b \approx 3.8$ (dashed line), and for RGF (3nn) (solid line). Kwant 3nn displayed the same trend as the corresponding nn.}\label{fig:cputime_lll}
\end{figure}

Let us begin considering a nearest neighbor (nn) tight-binding Hamiltonian in a two-dimension square-lattice of length $L$ and width $W$ in number of sites. We take $W'$ as the width of the leads (see inset of Fig.~\ref{fig:cputime_lll}a). We set $E=0$. In this case, we recall that for semi-infinite square lattice leads the number of open channels at the left and right leads $N_L=N_R=W'$. This model stems for instance from a finite-difference discretization of the Schr\"odinger equation of a mesoscopic two-dimensional electron gas (2DEG) \cite{Ferry1997,Datta1995}. For this model, the optimal partition of the RGF consists of $L$ partitions (slices) with $W$ sites each. 

\begin{figure}
    \centering
      \includegraphics[width=10cm]{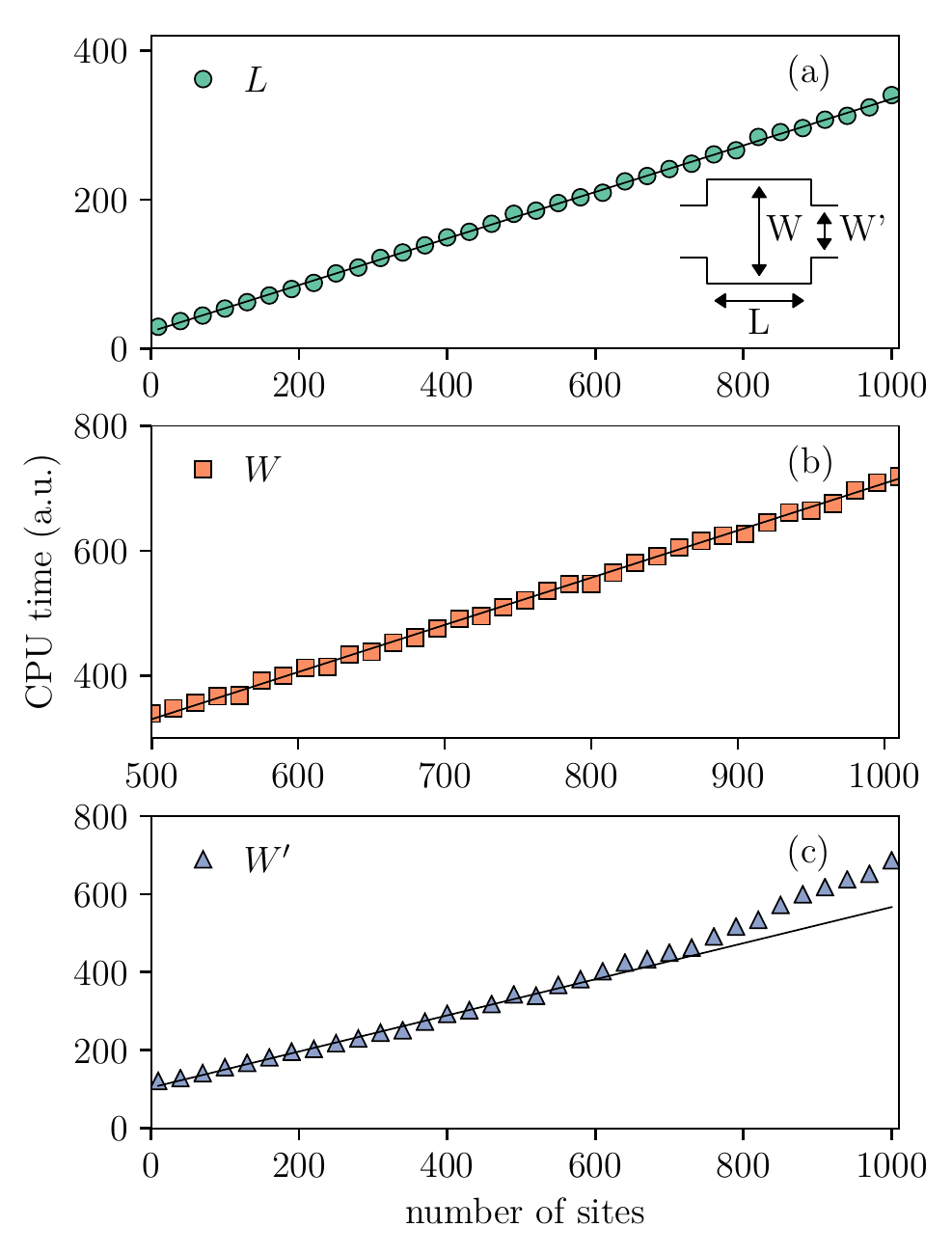}
    \caption{CPU time for the computation of the conductance as a function for a square lattice system of (a) length $L$ 
    (for $W$ and $W'$ fixed), (b) width $W$ (for $L$ and $W'$ fixed), 
    and (c) lead width $W'$ (for $L$ and $W$ constant). Since $E=0$, $W'=N_R = N_L$. Solid lines indicate linear fittings.}
    \label{fig:cputime_vs_sites}
\end{figure}

Figure \ref{fig:cputime_vs_sites} gives the CPU time (in arbitrary units) necessary to compute the conductance of the system, Eq.~\eqref{eq:landauer-buttiker}, as a function of $L$, $W$, and $W'$. It should be emphasized that in both implementations, the linear algebra calculations are coded in lower level programming languages, making this comparison possible.

As discussed in Sec.~\ref{sec:WFM} one has to solve $N_P$ times the sparse linear system of dimension $N_S+M_T$, Eq.(\ref{finalall}). Since the number of operations to solve a sparse system scales as $\mathcal{O}(N)$ \cite{Golub1996} and here $M_T = 2W'$, the WFM is expected scale as $(LW+ 2 W')W'$. Figure \ref{fig:cputime_vs_sites}a to \ref{fig:cputime_vs_sites}c verify that this conjecture is indeed correct. As a consequence, the performance of the WFM method is much better than previously believed \cite{Groth2014} for a realistic model of a nanostructure, $M_T=2W' \ll W$. 

Let us now examine a situation where $W=W'$. Figure \ref{fig:cputime_lll} clearly shows that the CPU time of the RGF (nn) method scales with $L^4$, as expected by the matrix multiplication and diagonalization operations involved. In distinction, the WFM shows a much better CPU performance scaling as $L^3$ (here $L=W=W'$). However, the overall pre-factor is typically large, making the method clearly advantageous 
only for $W\gtrsim 10^2$ sites.

We use this setting to investigate the efficiency of the WFM method when dealing with tight-binding Hamiltonians that consider hopping matrix elements beyond nearest-neighbor sites. This is the case in tight-binding models based on Wannier wave functions \cite{Calzolari2004}, that are very practical and accurate tools to model large scale disorder systems. Let us consider a square lattice tight-binding Hamiltonian with up to the $3$-rd nearest-neighbor (3nn) hopping terms. Since for the RGF method, only neighboring partitions should be connected in this model, one has to double the size of each slice, $W\to 2W$, reducing the total number of slices by half $L\to L/2$. Hence the CPU time grows by a factor of $4$ (solid line of Fig.~\ref{fig:cputime_lll}). In Fig.~\ref{fig:cputime_lll} we show that Kwant is practically insensitive to the coordination number of the lattice model, which represents a huge advantage over RGF.

\begin{figure}
 \centering
   \includegraphics[width=10cm]{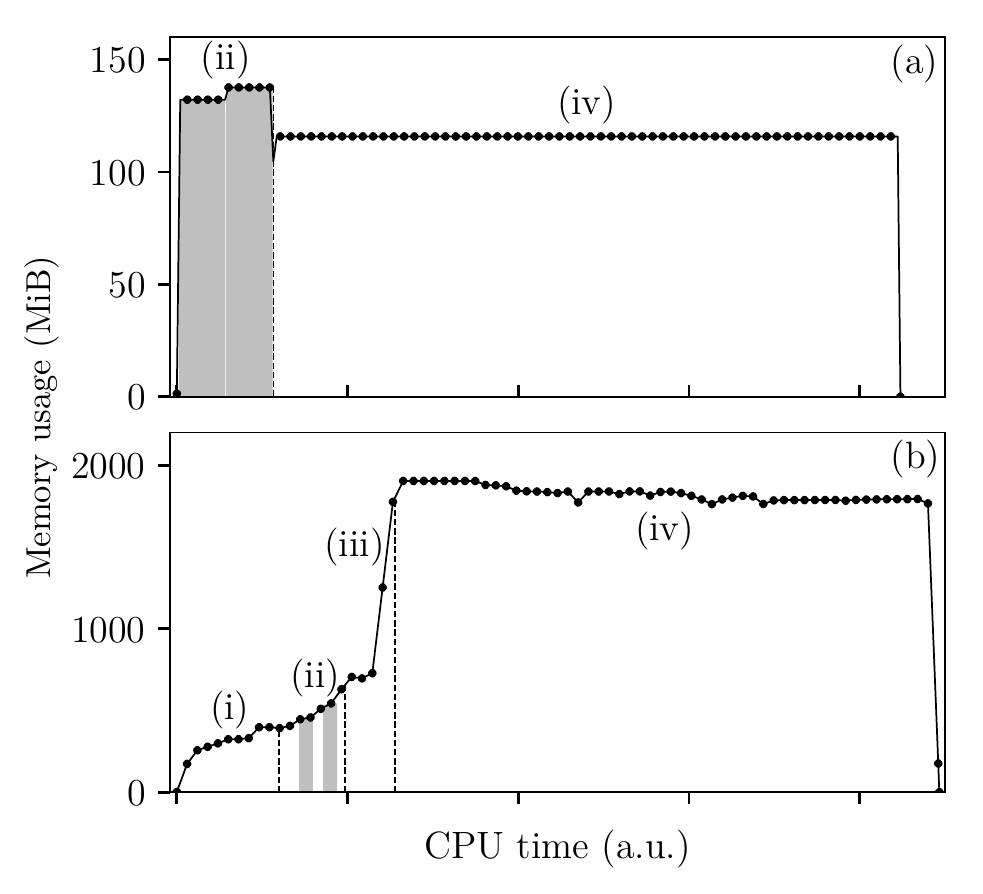}
   \caption{ Memory usage as a function of processing time in the calculation of the conductance  for a nearest-neighbor tight-binding 
   model of square lattice of dimensions $L=1000$ and $W=600$ for (a) the RGF and (b) the Kwant method. The different stages of the 
   computation are indicated by (i) to (iv), see main text.}
 \label{fig:memory_profile}
\end{figure}

Let us now analyze the memory usage of both methods. As already pointed out in Ref. \cite{Groth2014}, the memory usage in Kwant can be ten times larger than an RGF implementation which is a problem for computation of transport properties in large systems. In what follows we study this issue in more detail, examining the intermediate processes, such as the leads eigenmodes calculation, the linear system construction and factorization, and the solving stage, regions (ii)-(iv) of Fig.~\ref{fig:memory_profile}, respectively. This stage-by-stage information of the memory usage gives a clear view of the method advantages and bottlenecks.

Figure \ref{fig:memory_profile} shows the memory usage in a conductance calculation for both the WFM and the RGF implementations. A huge difference can be noted between the maximum memory used for each method. In Kwant, a preliminary time is spent in reading the input parameters, stage (i), which is negligible in the RGF Fortran 90 implementation and it is not displayed. The next stage in both methods, indicated by (ii) in Fig. \ref{fig:memory_profile}, is related to the computation of the lead contribution, namely, the lead surface Green's function in the RGF \cite{Datta1995} and the eigenmode diagonalization in WFM. In both methods, this is done twice for our two-probe model and $\Lambda-$times in general systems. Kwant spends an extra time in the factorization of the linear system, Eq.~(\ref{finalfull}).

At the solving stage, indicated by (iv) in Fig. \ref{fig:memory_profile}, we observe that Kwant requires one order of magnitude more memory than the RGF method. This is the only feature where the RGF outperforms the WFM methods. We note however that WFM approach allows for the computation of local operators (such as local currents and LDOS) with no significant additional cost, which is not the case for the RGF method.

Both methods are very robust and accurate. In our extensive tests, the computed conductances agree within the numerical precision. Even in the cases where the Green's function regularization factor $\eta$ is known to require a special choice in RGF, like transmission by evanescent modes in graphene \cite{Lima2016}, the WFM method gives reliable results without any particular adjustment.

\subsection{Application: Graphene Hall Bar}
\label{sec:graphene}

Let us now show the results of the WFM method for the transport properties in a realistic size graphene Hall bar in the quantum Hall regime \cite{Goerbig2011}. Despite the importance of such class of systems, few numerical studies have addressed the longitudinal and transverse resistances in Hall bar geometries due to the lack of an efficient multi-terminal electronic transport code.

\begin{figure}[ht!]
	\centering
	\includegraphics[width=16cm]{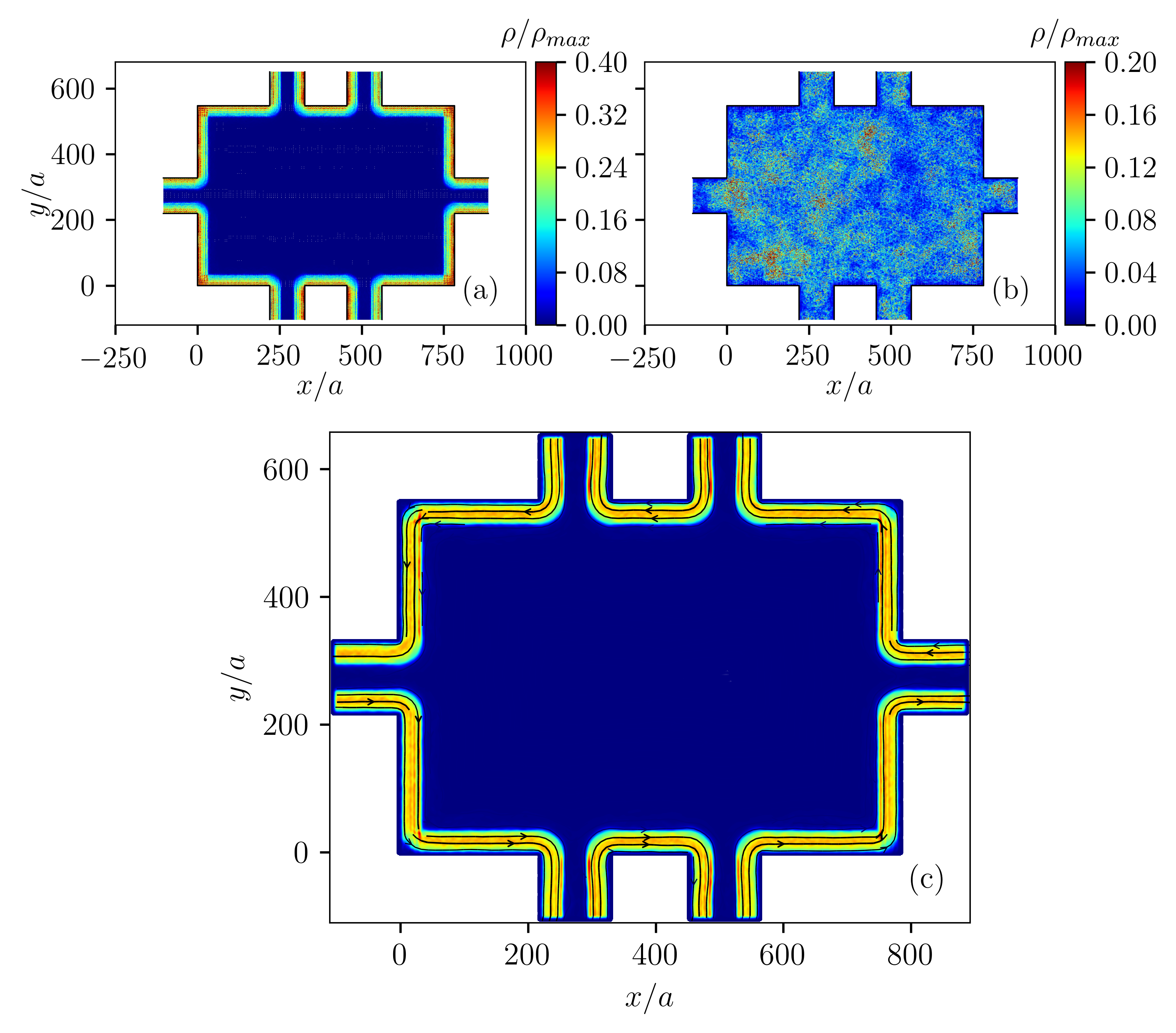}
	\caption{Local density of states at $E_F=0.5t$ and a magnetic flux of (a) $\phi = 0.004\phi_0$ (plateau state) and (b) $\phi =0.008\phi_0$ (transition state). (c) Local current at $E_F=0.5t$ and $\phi =0.004\phi_0$ (plateau state).}
	\label{fig:qhe}
\end{figure}

\begin{figure}[ht!]
	\centering
	\includegraphics[width=10cm]{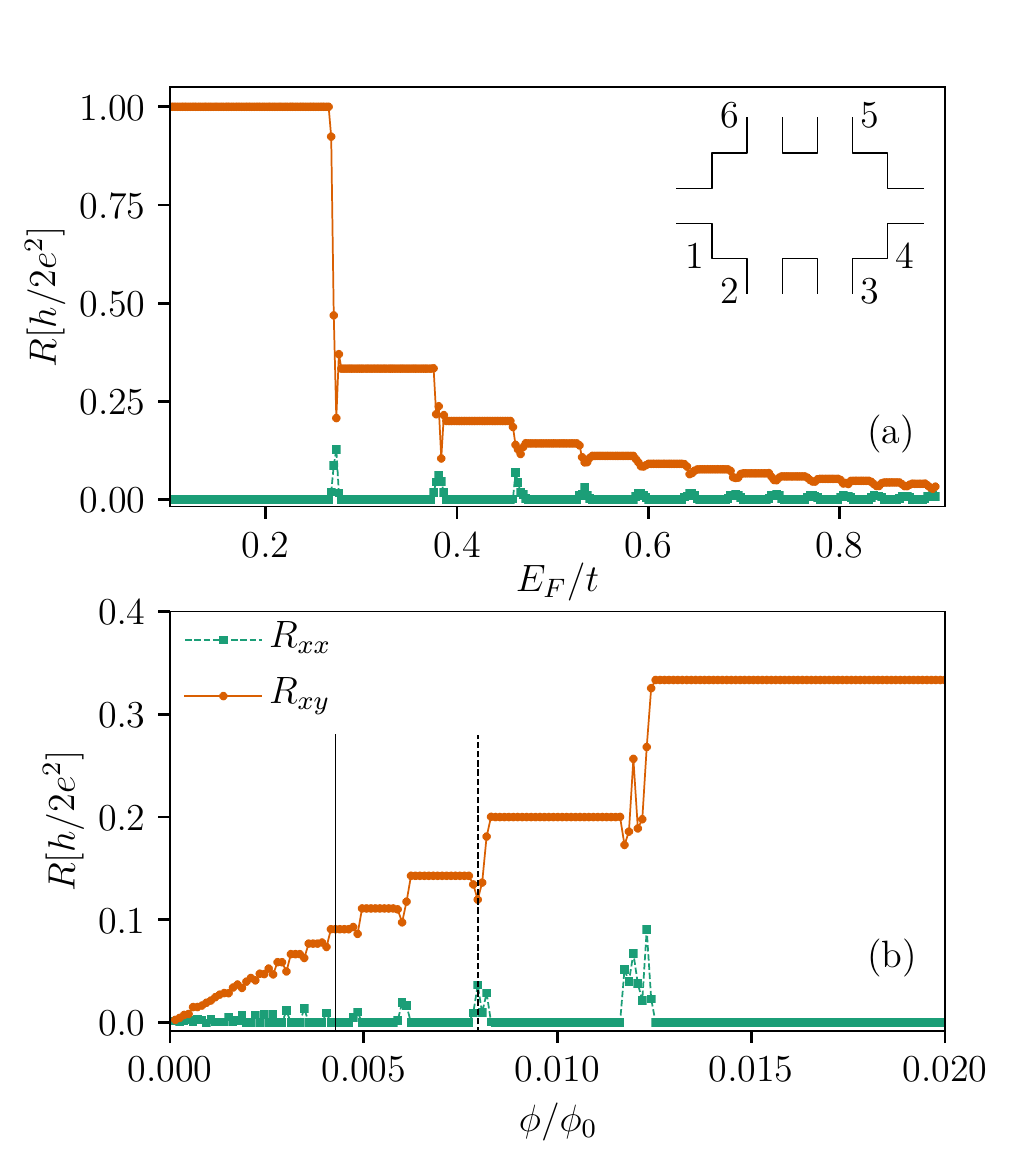}
	\caption{Graphene Hall bar longitudinal $R_{xx}$ and transverse $R_{xy}$ resistance for a single disorder realization 
	($10^6$ atoms and $T=0$) as a function of (a) $E_F/t$ for $\phi/\phi_0=0.007$ and (b) $\phi/\phi_0$ for $E_F/t=0.5$.}
	\label{fig:resistances_qhe}
\end{figure}

We consider a graphene sample with $\sim 10^6$ atoms in a Hall bar geometry (inset of Fig.~\ref{fig:resistances_qhe}). The graphene tight-binding Hamiltonian \cite{Goerbig2011} is 
$H=-\sum_{\langle i,j \rangle}(t_{ij}\ket{i}\bra{j}+{\rm H.c.})+\sum_i\epsilon_i \ket{i}\bra{i}$, where the sums run over the sites of a honeycomb lattice and $\langle\cdots\rangle$ restricts the pairs of sits to nearest-neighbors. The model includes a local (Anderson) scalar disorder by randomly choosing $\epsilon_i$ from an uniform distribution $[-\delta W,\delta W]$, where $\delta W=0.08t$.

The magnetic field ${\bf B} = B {\bf e}_z$ perpendicular to the graphene sheet is accounted for by Peierls substitution, namely, by taking $\phi_{ij}=\frac{e}{\hbar}\int_{{\bf r}_i}^{{\bf r}_j}{\bf A}({\bf r})\cdot d{\bf r}$ and ${\bf A}=B\left(-\beta y, (1-\beta) x,0 \right)$. The gauge $0\leq\beta\leq 1$ is conveniently chosen according to the orientation of the leads. Since in our Hall bar we consider leads along both the $x$ and $y$ directions, we avoid discontinuities in the magnetic field by smoothly varying the vector potential according to Ref.~\cite{Baranger1989,Power2017}.

We calculate the system longitudinal and transverse resistances using the Landauer-B\"uttiker formula, Eq. (\ref{eq:landauer-buttiker}), considering the terminals $\alpha = 2,3,5,6$ as voltage probes, that is, $I_2=I_3=I_5=I_6=0$ (see inset of Fig.~\ref{fig:resistances_qhe}b). In this setting, $I_1=-I_4=I$. Hence, $R_{xx}=(V_2-V_3)/I_1$ and $R_{xy}=(V_3-V_5)/I_1$. Our results for a single disorder realization correspond to typical quantum Hall resistance curves for graphene samples \cite{Goerbig2011}. Figure \ref{fig:resistances_qhe} shows quantized Hall plateaus at $R_{xy} = \frac{h}{2e^2}\frac{1}{2n+1}$ for integer values of $n$ and zero longitudinal resistance $R_{xx}$ at the $R_{xy}$ plateaus.

We also calculate the local density of states \cite{Groth2014}. Figures \ref{fig:qhe}a and \ref{fig:qhe}b show localized edge states along the sample in the plateau region and a delocalized state in the transition region between two plateaus (solid and dashed vertical lines of Fig.~\ref{fig:resistances_qhe}, respectively). As expected, the local current \cite{Groth2014} show quantum Hall edge states, see Fig.~\ref{fig:qhe}c.

The CPU time required for the resistance calculations is of the order of $80$ seconds for a single disorder realization and a single energy value in one core of an Intel\textregistered~Xeon\textregistered~X5650 processor. As discussed, such fast computation time in WFM relies on the $W'\ll W$ condition.

\section{Conclusion}
\label{sec:conclusion}

We have reviewed the underlying theory of the (WFM) method applied to a tight-binding (finite element) Hamiltonian used to model the transport properties of mesoscopic systems. Our analysis revealed that the WFM method is computationally far superior than previously expected \cite{Groth2014}.

We numerically verify ou predictions in a number of settings, benchmarking the CPU time, memory usage and precision of the WFM versus the RGF method.

To illustrate the power of the method we calculate the longitudinal and transverse resistance of a realistic-sized disordered graphene sheet in the quantum Hall regime. We consider a sample patterned in a Hall bar geometry, corresponding to a multi-terminal setting difficult to treat with other numerical approaches. 

We conclude mentioning that the WFM method allows for a straightforward generalization for multi-terminal systems with  nontrivial sample geometries, while the RGF approach resorts on ingenious schemes to deal with such situations \cite{Kazymyrenko2008,Lima2018}. In addition, the Kwant package also offers a set of implementation tools to facilitate the study of a wide range of settings, such as multi-orbital atomic states, general lattice connectivity and geometry, to name a few.

\section*{Acknowledgments}
We thank Xavier Waintal and Bruno A. D. Marques for useful discussions.
We acknowledge the financial support of the Brazilian funding agencies CNPq, CAPES, and FAPERJ.

\appendix

\section{Surface Green's function for a 1D chain}\label{app:surfaceGF}

In this Section we explicitly derive the Green's function of the first site in a semi-infinite linear chain of 
atoms using Eq.~(\ref{surfaceGF}). We consider a system with one orbital per atom, where the hopping 
matrices are numbers given by $H_T=\epsilon_0$ and $V_T=-t$. We apply the definition of the leads 
eigenstates in Eq.~(\ref{chi_lambda}) into Eq.~(\ref{lead_equation}) to find 
\begin{align}
	E = \epsilon_0 - t(\lambda^{-1}+\lambda),
	\label{dispersion1dchain}
\end{align}
whose solutions are
\begin{align}
	\lambda_1^\pm = -\left(\frac{E-\epsilon_0}{2t}\right) \pm i \sqrt{1-\left(\frac{E-\epsilon_0}{2t}\right)^2}.
\end{align}
Notice that if we substitute $\lambda = e^{ika}$ into Eq.~(\ref{dispersion1dchain}) we recover the well known 
dispersion relation for the 1D chain. 

The eigenstates corresponding to $\lambda_1^\pm$ are $\chi_1^\pm=1$. The probability current $j_1$, given 
by Eq.~(\ref{current}), reads
\begin{align}
	j_1^\pm = 2\frac{t}{\hbar} \text{Im}(\lambda_1^\pm) = \pm 2\frac{t}{\hbar} \sqrt{1-\left(\frac{E-\epsilon_0}{2t}\right)^2}.
\end{align}
Thus, there is one state propagating forwards ($j_1^+>0$) and one propagating backwards ($j_1^-<0$), resulting in $N_p=1$.

Since the dual vectors are simply $\tilde\chi_1^\pm=1$, Eq.~(\ref{Fdefinition}) leads to $F_- = \lambda_1^+$.
Therefore, the surface Green's function $G_T$ in Eq.~(\ref{surfaceGF}) reads
\begin{align}
	G_T = \left(E-\epsilon_0+t\lambda_1^+\right)^{-1} 
	    = \left(\frac{E-\epsilon_0}{2t}\right) - i \sqrt{1-\left(\frac{E-\epsilon_0}{2t}\right)^2}.
	\label{surfaceGF1d}
\end{align}

\bibliographystyle{model1-num-names}
\bibliography{wfm_method}

\end{document}